\RequirePackage{fix-cm}
\documentclass[twocolumn,epjc3,final]{svjour3}  
\smartqed  % flush right qed marks, e.g. at end of proof
 \usepackage{mathrsfs}
\usepackage{amsmath, txfonts}
\usepackage{graphicx,bm,booktabs,bm}
\usepackage{longtable,lscape}
\usepackage{ulem} 
\usepackage{overpic,multirow,color}
\definecolor{cover}{rgb}{0.77,0.87,0.88}
\definecolor{blueone}{rgb}{0.1,0.1,.7}
\definecolor{citec}{rgb}{0.14,0.47,0.09}
\definecolor{two}{rgb}{0.0,0.5,0.}
\definecolor{three}{rgb}{.5,.1,0.15}
\usepackage[bookmarks=true,bookmarksopen=false,plainpages=false,breaklinks=true,
   bookmarksnumbered=true,hypertexnames=false,
   filecolor=blue,urlcolor=three,menucolor=three,
   linkcolor=three,citecolor=blueone, colorlinks,
   anchorcolor=blue,runcolor=pink,frenchlinks=red
   pdfstartview=FitH,pdftitle=title,%
   pdfauthor=author]{hyperref}
              \allowdisplaybreaks[4]

\graphicspath{{Figures/}} %

\journalname{Eur. Phys. J. C}
\begin{document}
\title{$Z_c(3900)/Z_c(3885)$ as a  virtual  state from $\pi J/\psi-\bar{D}^*D$ interaction}
\author{Jun He\thanksref{e1,addr1}
\and Dian-Yong Chen\thanksref{addr2}
}                     % Do not remove
\thankstext{e1}{Corresponding author: junhe@njnu.edu.cn}
\institute{Department of  Physics and Institute of Theoretical Physics, Nanjing Normal University,
Nanjing 210097, China\label{addr1}
\and
School of Physics, Southeast University, Nanjing 210094,  China\label{addr2}
}

\date{Received: date / Revised version: date}
\maketitle

\abstract{
In this work, we study the $J/\psi\pi$ and $\bar{D}^*D$  invariant mass spectra of the $Y(4260)$ decay to find out  the origin of the $Z_c(3900)$ and $Z_c(3885)$ structures.  The $J/\psi \pi-\bar{D}^{*}D$ interaction is studied in a coupled-channel quasipotential Bethe-Saltpeter equation approach, and embedded to the $Y(4260)$ decay process to reproduce both  $J/\psi\pi^-$ and   $D^{*-}D^0$ invariant mass spectra observed at BESIII simultaneously.  It is found out that a virtual state at energy about 3870 MeV is produced from the interaction when both invariant mass  spectra are comparable with the experiment. The  results support that both $Z_c(3900)$ and $Z_c(3885)$ have the same origin, that is, a virtual  state from  $J/\psi \pi-\bar{D}^*{D}$ interaction, in which the $\bar{D}^*{D}$ interaction is more important and the coupling between   $\bar{D}^*{D}$  and $J/\psi\pi$ channels plays a minor role. 
} %end of abstract

\section{Introduction}\label{sec1}

In  recent years, many exotic resonance structures were observed near the threshold of two hadrons, which are difficult to  put into the conventional quark model.  The exotic resonance structures near the $\bar{D}^*D$ threshold (in this work we will remark hidden charmed system with a vector $D^*/\bar{D}^*$ meson and a pseudoscalar $\bar{D}/D$ meson as $\bar{D}^*D$ if the explicit is not necessary) are good examples of such phenomena.  The first $XYZ$ particle, $X(3872)$ is almost on the $\bar{D}^*D$ threshold, which was interpreted as a  $\bar{D}^\ast D $ hadronic molecular state immediately after its observation \cite{Choi:2003ue,Tornqvist:2004qy}.
Later, an isovector resonant structure named $Z_c^\pm(3900)$ was observed in 2013 at BESIII and Bell in the $J\psi \pi^{\pm}$ invariant mass spectrum of $e^+ e^- \to \pi^+ \pi^- J/\psi$ at $\sqrt{s}=4.26$ GeV \cite{Ablikim:2013mio,Liu:2013dau}, and further confirmed at CLEO-c in the same channel at $\sqrt{s}=4.17$ GeV \cite{Xiao:2013iha}. The observed $Z_c(3900)$ is also near the $\bar{D}^\ast D$ threshold, thus, it is natural to explain it as an isovector partner of $X(3872)$ in the $\bar{D}^\ast D$ molecular scenario and expected to be observed in the $
(\bar{D}^\ast D)_{I=1}$ channel.  It was confirmed by the observation of $Z_c(3885)$ in the $D\bar{D}^*$  invariant mass spectrum of $Y(4260)$ decay
in process $e^+e^-\to\pi^\pm (D\bar{D}^*)^\mp$~\cite{Ablikim:2013emm}.
Recently, the neutral
partners  of $Z^\pm_c(3900)$ and $Z_c^\pm(3885)$ were also observed at BESIII~\cite{Ablikim:2015tbp,Ablikim:2015gda}. Besides, the spin parity of these state has been determined as $J^P=1^+$ by a partial wave analysis~\cite{Collaboration:2017njt}.  

Thanks to  the experiments at BESIII, Belle and CLEO-c, the isovector $Z_c(3900)/Z_c(3885)$ has been  established.
Since the $Z_c^\pm(3900)/Z_c(3885)$ carries a charge,
it cannot be explained as  a
$c\bar{c}$ state, which must be neutral.  After the observation at BESIII, many interpretations of the origin of $Z_c(3900)/Z_c(3885)$ have been proposed, which includes the hadronic molecular state~\cite{Guo:2013sya,Wang:2013cya,Wilbring:2013cha,He:2014nya,Chen:2015igx},  tetraquark state ~\cite{Braaten:2013boa,Dias:2013xfa,Wang:2013vex}, initial-single-pion-emission mechanism~\cite{Chen:2011xk,Chen:2013coa}, cusp effect from triangle singularity~\cite{Liu:2013vfa}.
Due to its closeness to the $\bar{D}^*D$ threshold,
the hadronic molecular state is an important picture to explain the  $Z_c(3900)/Z_c(3885)$ structure. In Ref.~\cite{Sun:2011uh,Sun:2012zzd}, the  $\bar{D}^*D$ interaction as well as the $\bar{B}^*B$ interaction was studied in a one-boson-exchange model.  No bound state was found with the light-meson exchange.  In the chiral unitary approach, the contribution of exchange of heavy meson was included into  the  $\bar{D}^*D$ interaction, which provides   attraction strong enough to produce a bound state~\cite{Aceti:2014uea}. The importance of the heavy-meson exchange was confirmed by a further study in the one-boson-exchange model combined with a quasipotential Bethe-Salpeter equation approach~\cite{He:2015mja}.
 
The early studies in the hadronic molecular picture focus on how to produce a bound state corresponding to the $Z_c(3900)/Z_c(3885)$ from  the $\bar{D}^*D$ interaction. It is interesting to study if the bound state obtained in those studies can reproduce the original experimental data of the invariant mass spectra.  In Ref.~\cite{Aceti:2014uea}, the invariant mass spectra were studied  in a chiral unitary approach while the Breit-Wigner form with  mass and width obtained from the interaction were adopted. In Ref.~\cite{Chen:2013coa} the invariant mass spectra were studied in an initial-single-pion-emission mechanism, where the $\bar{D}^*D$ rescattering were not considered.  Besides, when this study was done, only $\pi J/\psi$ invariant mass spectrum was available. In Refs.~\cite{Albaladejo:2015lob,Pilloni:2016obd,Zhou:2015jta,Gong:2016hlt}, the mass invariant mass spectra was explicitly studied and fitted, and the poles corresponding to the $Z_c(3900)/Z_c(3885)$ are extracted, especially in Ref.~\cite{Albaladejo:2015lob} the analysis suggested the $Z_c(3885)/Z_c(3900)$ maybe originate from a virtual state. However, in theses studies, all coupling constants of the interactions were chosen as free parameters.  
A recent lattice work suggests that the off-diagonal $\pi J/\psi-\bar{D}D^*$ coupling is more important than the $ \bar{D}D^*$ interaction, and a semiphenomenological  analysis was adopted to study the invariant mass spectrum~\cite{Ikeda:2016zwx}.   In their comparison with data, two general free parameters were adopted to $\pi J/\psi$ and $\bar{D}D^*$ channel, respectively, which smeared an important experiment results about the relative magnitudes of decays in these two channels. BESIII reported that the decay width of $Z_c(3885)$ in $D\bar{D}^*$  is still much larger than that of $Z_c(3900)$ in $J/\psi \pi$ channel with a ratio $6.2\pm1.1\pm2.7$, though which is much smaller than conventional charmonium states above the open charm threshold~\cite{Ablikim:2013xfr}. 

In this work, we try to reproduce  both line shapes and relative magnitudes of the  $\pi J/\psi$ and $\bar{D}^*D$ invariant mass spectra, simultaneously. It is performed by studying the $Y(4260)$ decay with reacattering of $\pi J/\psi-\bar{D}^*D$, which is calculated in a quasipotential Bethe-Salpeter equation approach.  In the calculation, the interaction is constructed with the Lagrangians from the heavy-quark effective theory.  And in this work, we only consider the system with negative charge, the positive and neutral cases are analogous due to the SU(3) symmetry.

In the next section, the formalism adopted to calculate the three-body decay of the $Y(4260)$ in the current work is presented. The interaction potential is constructed with an effective Lagrangian and the quasipotential Bethe-salpeter equation will be introduced briefly. The numerical results are given in Section \ref{sec3}. A brief summary  is given in the last section.

\section{Formalism of three-body decay}\label{sec2}

\begin{figure}[h!]\begin{center}
\includegraphics[bb=78 600 560 780,clip,scale=0.51]{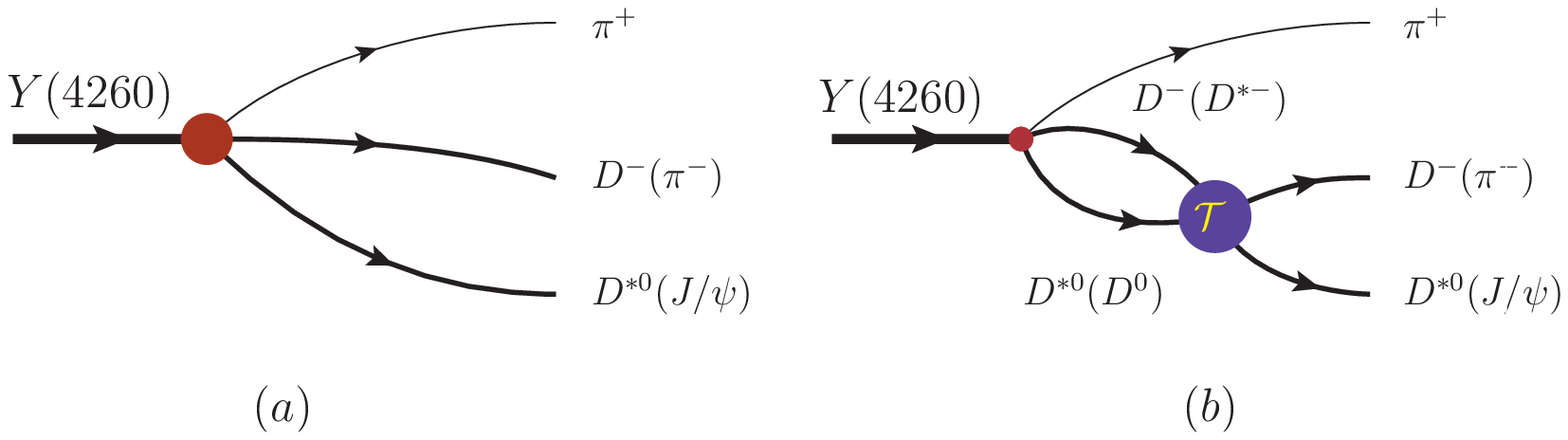}
\end{center}\caption{The diagrams for the $Y(4260)$ decay. Diragrams (a) and (b) are for background and rescattering  contribution.\label{Fig: Ydecay}}
\end{figure}

The $Z_c(3900)/Z_c(3885)$ resonance structure were observed at the $e^+e^-\to Y(4260) \to\pi Z_c \to \pi(\pi J/\psi/\bar{D}^*{D})$ process at BESIII. The internal structure of the $Y(4260)$ is still in the debate. To avoid the complexity, we adopt a phenomenological vertex $Y(4260)\to \pi (\bar{D}^*{D})$.  The effect of the $e^-e^-\to Y(4260)$ is also absorbed into this vertex.  Hence, to study the invariant mass spectrum, we  consider the  three-body-decay diagram  in Fig.~\ref{Fig: Ydecay}.   In this work, we focus on the invariant mass spectrum near the $\bar{D}D^*$ threshold and the $J/\psi\pi$ interaction is suppressed by the OZI rule, The explicit calculation in our model also suggests that the intermediate $\pi J/\psi$ channel will be suppressed seriously. Hence, we only consider the intermediate $D^-D^{*0}$ and $D^{*-}D^0$  channels in the loop between direct decay vertex and the rescattering from beginning, which was also adopted in Ref.~\cite{Gong:2016hlt}. When calculating the rescattering amplitude ${\cal T}$, the  $\pi^-J/\psi -D^-D^{*0}-D^{*-}D^0$ interaction is considered.

With the decay amplitude ${\cal M}$ the invariant of the mass spectrum can be obtained from  the differential decay width of $Y(4260)$ as
\begin{align}
d\Gamma&=\frac{1}{2M}|{\cal M}|^2 d\Phi,
\end{align}
where the $M$ is the mass of the $Y(4260)$ and 
phase space can be written as
\begin{align}
d\Phi
&=\frac{1}{(2\pi)^5}\frac{\breve{\rm p}_1{\rm p}^{cm}_3}{M}d\Omega_1d\Omega^{cm}_3dM_{23},
\end{align}
Here $cm$ means the center of mass frame of particles 2  and 3. The explicit deduction  is given in ~\ref{Sec: phasespace}.

\begin{figure}[h!]\begin{center}
\includegraphics[bb=60 710 560 780,clip,scale=0.5]{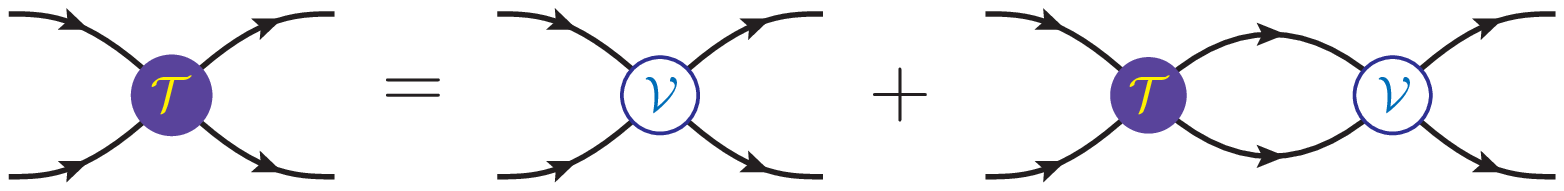}
\end{center}\caption{The diagrams for the Bethe-Salpeter equation. \label{Fig: BS}}
\end{figure}

The key to study the decay amplitude is to write rescattering amplitude ${\cal T}$ of the $J/\psi^--D^-D^{*0}-D^{*-}D^0$ interaction, which can be obtained with the help of the Bethe-Salpeter equation as shown in Fig.~\ref{Fig: BS}.

To avoid  difficulty of solving a four-dimensional equation in the Minkowski space,  with quasipotential approximation, the Bethe-Salpeter equation is often reduced to a three-dimensional equation, which can be further reduced to a one-dimensional equation by  partial wave decomposition. In this work, the OBE interaction will be adopted.  The off-shellness  of two constituent hadrons should be kept to avoid the unphysical singularity below the threshold. The covariant spectator theory, in which  the heavier constituent is put on shell~\cite{Gross:1991pm,VanOrden:1995eg,Gross:1999pd,He:2013oma,He:2015yva},  will be adopted in our study of the  $J/\psi^--D^-D^{*0}-D^{*-}D^0$ interaction. Such treatment was explained explicitly in the appendices of Ref.~\cite{He:2015mja} and has been applied to studied the   X(3250), the $Z_c(3900)$ and the LHCb pentaquarks and its strange partners~\cite{He:2014nya,He:2012zd,He:2015cea,He:2017aps}.  The partial-wave Bethe-Salpeter equation with fixed
spin parity $J^P$ of system is written as~\cite{He:2015mja}
\begin{align}
i{\cal T}^{J^P}_{\lambda'_2\lambda'_3,\lambda_2\lambda_3}({\rm p}',{\rm p})
&=i{\cal V}^{J^P}_{\lambda'_2\lambda'_3,\lambda_2\lambda_3}({\rm p}',{\rm
p})+\sum_{\lambda''_2\lambda''_3\ge0}\int\frac{{\rm
p}''^2d{\rm p}''}{(2\pi)^3}\nonumber\\
&\cdot
i{\cal V}^{J^P}_{\lambda'_2\lambda'_3,\lambda''_2\lambda''_3}({\rm p}',{\rm p}'')
G_0({\rm p}'')i{\cal T}^{J^P}_{\lambda''_2\lambda''_3,\lambda_2\lambda_3}({\rm p}'',{\rm
p}),\quad\quad \label{Eq: BS_PWA}
\end{align}
with the
reduced propagator written down in the center-of-mass frame with $P=(M,{\bm 0})$ as
\begin{align}
	G_0&=\frac{\delta^+(p''^{~2}_h-m_h^{2})}{p''^{~2}_l-m_l^{2}}
	\nonumber\\&=\frac{\delta^+(p''^{0}_h-E_h({\bm p}''))}{2E_h({\bm p''})[(W-E_h({\bm
p}''))^2-E_l^{2}({\bm p}'')]}.
\end{align}
Here the heavier particle (remarked with $h$)  is put on shell, which has  $p''^0_h=E_{h}({\rm p}'')=\sqrt{
m_{h}^{~2}+\rm p''^2}$. The $p''^0_l$ for the lighter particle (remarked as $l$) is then $W-E_{h}({\rm p}'')$. Here and hereafter we will adopt  a definition ${\rm p}=|{\bm p}|$. And the momentum of particle 2  ${\bm p}''_2=-{\bm
p}''$ and the momentum of particle 3 ${\bm p}''_3={\bm
p}''$. 
The potential kernel ${\cal V}^{J^P}_{\lambda'_2\lambda'_3\lambda_2\lambda_3}$  with spin-parity $J^P$ is defined   as
\begin{eqnarray}
i{\cal V}_{\lambda'_2\lambda'_3\lambda_2\lambda_3}^{J^P}({\rm p}',{\rm p})
&=&2\pi\int d\cos\theta
~[d^{J}_{\lambda_{32}\lambda'_{32}}(\theta)
i{\cal V}_{\lambda'_2\lambda'_3\lambda_2\lambda_3}({\bm p}',{\bm p})\nonumber\\
&+&\eta d^{J}_{-\lambda_{32}\lambda'_{32}}(\theta)
i{\cal V}_{\lambda'_2\lambda'_3-\lambda_2-\lambda_3}({\bm p}',{\bm p})],
\end{eqnarray}
where $\lambda_{32}=\lambda_3-\lambda_2$ and $\eta=PP_2P_3(-1)^{J-J_2-J_3}$ with $J_{(2,3)}$ and $P_{(2,3)}$ being the spin and parity of  constituent 2 or 3. Without loss of  generality  the initial and final relative momenta are chosen as ${\bm p}=(0,0,{\rm p})$  and ${\bm p}'=({\rm p}'\sin\theta,0,{\rm p}'\cos\theta)$, and the $d^J_{\lambda\lambda'}(\theta)$ is the Wigner d-matrix.

In most cases, the integral in Eq.~(\ref{Eq: BS_PWA}) is non-convergent. In this work an exponential regularization is introduced by
a replacement of the propagator as
 \begin{eqnarray}
 G_0({\rm p})\to G_0({\rm
 p})\left[e^{-(p''^2_l-m_l^2)^2/\Lambda^4}\right]^2.\label{Eq: FFG}
 \end{eqnarray}
We would like to remind that the regularization of  heavier particle  vanishes because it is put onshell in the quasipotential approximation adopted. With the regularization, the contributions at large momentum ${\rm p}''$ is suppressed heavily at the energies higher than 2 GeV~\cite{He:2016pfa}, which guarantees the  convergence of the integral. if we multiply exponential factor on both sides of the equation~(\ref{Eq: BS_PWA}), it can be found that the regularization factor can be seen as a form factor introduced due to the off-shell effect of particle 1 in a form of $e^{-(k^2-m^2)^2/\Lambda^4}$. 
The interested reader is referred to Ref.~\cite{He:2015mja} for
further information about the regularization. 

The write the amplitude of three-body decay of the $Y(4260)$, we adopt an effective  Lagrangian for the $Y\to \pi DD^*$ as,
\begin{align}
{\cal L}_{Y\to\pi DD^*}&=g_{Y\to\pi DD^*}Y^\mu(D{\bm \tau}\cdot{\bm \pi} \bar{D}^*_\mu+{D}^*_\mu{\bm \tau}\cdot{\bm \pi}\bar{D}),
\end{align}
The partial-wave amplitudes with spin parity $J^P$ is
\begin{align}
{\cal A}^{J^P}_{\lambda_2,\lambda_3;\lambda}({\rm p}^{cm})&=\int d\Omega_3^{cm}[{\cal A}_{\lambda_2,\lambda_3;\lambda}(P,p^{cm}_2,p^{cm}_3)D^{J*}_{\lambda_R,\lambda_{32}}( \Omega_3^{cm})\nonumber\\
&\eta{\cal A}_{-\lambda_2,-\lambda_3;\lambda}(P,p^{cm}_2,p^{cm}_3)D^{J*}_{\lambda_R,-\lambda_{32}}( \Omega_3^{cm}).
\end{align}
With Lagrangian we adopted, only the  $J^P=0^+$ and $1^-$ partial wave survive as
\begin{align}
{\cal A}^{1^+}_{\lambda_2,\lambda_3;\lambda}
&=\frac{2}{N_1^2}(\delta_{\lambda_3\pm}+\frac{E^{cm}}{m}\delta_{\lambda_30})(\delta_{\lambda\pm}+\frac{P^{0cm}}{M}\delta_{\lambda0})D^1_{\lambda\lambda}(\Omega_1).\nonumber\\
{\cal A}^{0^-}_{\lambda_2,\lambda_3;\lambda}&=\frac{2}{N_0^2}\delta_{\lambda_30}\frac{{\rm p}^{cm}}{m}\delta_{\lambda0}\frac{{\rm P}^{cm}}{M^{cm}}.\label{Eq: Ad}
\end{align}

The total three-body decay with the rescattering is written as
\begin{align}
&{\cal M}^{Z}_{\lambda_2,\lambda_3;\lambda}(p_1,p_2,p_3)\nonumber\\&=\sum_{J\lambda_R}N_{J}D^{J*}_{\lambda_R,\lambda_{32}}( \Omega_3^{cm})\sum_{\lambda'_2\lambda'_3}\int \frac{{\rm p}'^{cm2}_3d{\rm p}'^{cm}_3}{(2\pi)^3}\nonumber\\ &\cdot ~i{\cal T}^J_{\lambda_2,\lambda_3;\lambda'_2,\lambda'_3}({\rm p}'^{cm}_3) G_0({\rm p}'^{cm}_3) {\cal A}^{J}_{\lambda'_2,\lambda'_3;\lambda}({\rm p}'^{cm}_3,\Omega_1).
\end{align}

The distribution can be obtained as
\begin{align}
{d\Gamma\over dM_{23}}&=\int \frac{1}{6M}\sum_{\lambda_2,\lambda_3;\lambda}|{\cal M}_{\lambda_2,\lambda_3;\lambda}|^2 \frac{1}{(2\pi)^5}\frac{\breve{\rm p}_1{\rm p}^{cm}_3}{M}d\Omega_1d\Omega^{cm}_3\nonumber\\
&=\frac{1}{6M}\frac{1}{(2\pi)^5}\frac{\breve{\rm p}_1{\rm p}^{cm}_3}{M}\sum_{\lambda_2,\lambda_3;\lambda;J}|\hat{\cal M}^J_{\lambda_2,\lambda_3;\lambda}(M_{23})|^2 .
\end{align}
Here the explicit form of ${\cal A}^{J}_{\lambda'_2,\lambda'_3;\lambda}({\rm p}'^{cm}_3,\Omega_1)$ with $J=0,1$ in Eq.~(\ref{Eq: Ad}) is applied. 

The distribution can be further rewritten with the partial wave amplitudes withe $J^P$ as
\begin{align}
{d\Gamma\over dM_{23}}&=\frac{1}{6M}\frac{1}{(2\pi)^5}\frac{\breve{\rm p}_1{\rm p}^{cm}_3}{M}\sum_{i\ge0;j\ge0;J^P}\frac{1}{N_J^2}|\hat{\cal M}^{J^P}_{i;j}(M_{23})|^2, 
\end{align}
with 
\begin{align}
\hat{\cal M}^{J^P}_{i;i}(M_{23})
&=\hat{\cal A}^{bk,J^P}_{j;i}(M_{23})+\sum_{k}\int \frac{d{\rm p}'^{cm}_3{\rm p}'^{cm2}_3}{(2\pi)^3}\nonumber\\
&i\hat{\cal T}^{J^P}_{j;k}({\rm p}'^{cm}_3,M_{23}) G_0({\rm p}'^{cm}_3) \hat{\cal A}^{J^P}_{k;i}({\rm p}'^{cm}_3,M_{23})
\end{align}
where $i$ and $j$ denote the independent $\lambda_{2,3}$ and $\lambda$, and the factors $f_{i=0}=1/\sqrt{2}$ and $f_{i\neq0}=1$ are inserted.  
In this work we introduce  parameterized background contribution  with 
\begin{align}
\hat{\cal A}^{bk,J^P}_{j;i}(M_{23})=c(M_{23}-M_{min})^a(M_{max}-M_{23})^b~\hat{\cal A}^{J^P}_{j;i}(M_{23}).
\end{align}
The parameters will be determined by comparing with experiment.

\section{Lagrangians and $\pi^-J/\psi-D^-D^{*0}/D^{*-}D^0$ interaction}

Now we need to construct the  the potential ${\cal V}$ of  the  $\pi^-J/\psi-D^-D^{*0}-D^{*-}D^0$ interaction to provide the rescattering amplitude ${\cal T}$.  In this work, we adopt the Lagrangians from the heavy quark effective theory. The effective Lagrangian of
the pseudoscalar mesons with heavy flavor mesons reads~\cite{Colangelo:2003sa,Casalbuoni:1996pg}
\begin{align}\label{eq:lag-p-exch}
  \mathcal{L}_{D^*D\mathbb{P}} &=
  -i\frac{2g\sqrt{m_Dm_{D^*}}}{f_\pi} (-D_bD^{*\dag}_{a\lambda}+D^*_{b\lambda}D^\dag_{a})\partial^\lambda{}\mathbb{P}_{ba}
 \nonumber\\
 & +i\frac{2g\sqrt{m_Dm_{D^*}}}{f_\pi}
  (-\tilde{D}^{*\dag}_{a\lambda}\tilde{D}_b+\tilde{D}^\dag_{a}\tilde{D}^*_{b\lambda})\partial^\lambda\mathbb{P}_{ab},\nonumber\\
    \mathcal{L}_{D^*D^*\mathbb{P}} &=
  \frac{g}{f_\pi}\epsilon_{\alpha\mu\nu\lambda}D^{*\mu}_b\overleftrightarrow{\partial}^\alpha D^{*\lambda\dag}_{a}\partial^\nu\mathbb{P}_{ab}-\frac{g}{f} \epsilon_{\alpha\mu\nu\lambda}\tilde{D}^{*\mu\dag}_a\overleftrightarrow{\partial}^\alpha D^{*\lambda}_{b}\partial^\nu\mathbb{D}_{ba},\nonumber\\
  \mathcal{L}_{D^*D\mathbb{V}} &=
  \sqrt{2}\lambda g_V\varepsilon_{\lambda\alpha\beta\mu}
  (-D^{*\mu\dag}_a\overleftrightarrow{\partial}^\lambda D_b
  +D^\dag_a\overleftrightarrow{\partial}^\lambda D_b^{*\mu})
  (\partial^\alpha{}\mathbb{V}^\beta)_{ba}\nonumber\\
  &+\sqrt{2}\lambda g_V\varepsilon_{\lambda\alpha\beta\mu}
  (-\tilde{D}^{*\mu\dag}_a\overleftrightarrow{\partial}^\lambda
  \tilde{D}_b  +\tilde{D}^\dag_a\overleftrightarrow{\partial}^\lambda
  \tilde{D}_b^{*\mu})(\partial^\alpha{}\mathbb{V}^\beta)_{ab},
\end{align}
with the octet pseudoscalar and nonet vector
meson matrices as
\begin{eqnarray}
\mathbb{P}&=&\left(\begin{array}{ccc}
\frac{\pi^{0}}{\sqrt{2}}+\frac{\eta}{\sqrt{6}}&\pi^{+}&K^{+}\\
\pi^{-}&-\frac{\pi^{0}}{\sqrt{2}}+\frac{\eta}{\sqrt{6}}&
K^{0}\\
K^- &\bar{K}^{0}&-\frac{2\eta}{\sqrt{6}}
\end{array}\right),\nonumber\\
\mathbb{V}&=&\left(\begin{array}{ccc}
\frac{\rho^{0}}{\sqrt{2}}+\frac{\omega}{\sqrt{2}}&\rho^{+}&K^{*+}\\
\rho^{-}&-\frac{\rho^{0}}{\sqrt{2}}+\frac{\omega}{\sqrt{2}}&
K^{*0}\\
K^{*-} &\bar{K}^{*0}&\phi
\end{array}\right).\label{vector}
\end{eqnarray}
which correpond to $(D^0,D^+,D_s^+)$
	and $(\bar{D}^0,D^-,D_s^-)$.

The effective Lagrangian of the vector mesons with heavy flavor
mesons reads
\begin{eqnarray}\label{eq:lag-v-exch}
	\mathcal{L}_{\mathcal{DD}\mathbb{V}} &=& -i\frac{\beta
	g_V}{\sqrt{2}} D_a^\dag\overleftrightarrow{\partial}^\mu D_b\mathbb{V}^\mu_{ba}
	+i\frac{\beta	g_V}{\sqrt{2}}\tilde{D}_a^\dag
	\overleftrightarrow{\partial}^\mu \tilde{D}_b\mathbb{V}^\mu_{ab},\nonumber\\
  \mathcal{L}_{\mathcal{D^*D^*}\mathbb{V}} &=& i\frac{\beta g_V}{\sqrt{2}}
  D_a^{*\dag}\overleftrightarrow{\partial}^\mu D^*_b\mathbb{V}^\mu_{ba}
 \nonumber\\& -&i2\sqrt{2}\lambda
  g_V m_{D^*}D^{*\mu}_bD^{*\nu\dag}_a(\partial_\mu\mathbb{V}_\nu-\partial_\nu\mathbb{V}_\mu)_{ba}
  \nonumber\\
  &-& i\frac{\beta
  g_V}{\sqrt{2}}\tilde{D}_a^{*\dag}\overleftrightarrow{\partial}^\mu
  \tilde{D}^*_b\mathbb{V}^\mu_{ab}
  \nonumber\\
  &-& i2\sqrt{2}\lambda  g_Vm_{D^*}\tilde{D}^{*\mu\dag}_a\tilde{D}^{*\nu}_b(\partial_\mu\mathbb{V}_\nu-\partial_\nu\mathbb{V}_\mu)_{ab}
,\nonumber\\
  \mathcal{L}_{\mathcal{DD}\sigma} &=&
  -2g_\sigma m_{D}D_a^\dag D_a\sigma -2g_\sigma m_{D}\tilde{D}_a^\dag \tilde{D}_a\sigma,\nonumber\\
  \mathcal{L}_{\mathcal{D^*D^*}\sigma} &=&
  2g_\sigma m_{D^*}D_a^{*\dag} D^*_a\sigma +2g_\sigma m_{D^*}\tilde{D}_a^{*\dag}
  \tilde{D}^*_a\sigma.
\end{eqnarray}
Here the parameters are determined as $g=0.59$, $\beta$=0.9, $\lambda$=0.56 GeV$^{-1}$, $g_V=5.8$ and $g_\sigma = g_\pi/(2\sqrt{6})$ with $g_\pi = 3.73$~\cite{Isola:2003fh,Falk:1992cx}.

The couplings of heavy-light charmed mesons to $J/\psi$ follow
form,
\begin{eqnarray}
	{\cal L}_{D^*\bar{D}^*J/\psi}&=&-ig_{D^*D^*\psi}
\big[\psi \cdot \bar{D}^*\overleftrightarrow{\partial}\cdot D^*  \nonumber\\&-&
\psi^\mu \bar D^* \cdot\overleftrightarrow{\partial}^\mu {D}^* +
\psi^\mu \bar{D}^*\cdot\overleftrightarrow{\partial} D^{*\mu} ) \big] ,\nonumber \\
{\cal L}_{D^*\bar{D}J/\psi}&=&
-g_{D^*D\psi} \,  \, \epsilon_{\beta \mu \alpha \tau}
\partial^\beta \psi^\mu (\bar{D}
\overleftrightarrow{\partial}^\tau D^{* \alpha}+\bar{D}^{* \alpha}
\overleftrightarrow{\partial}^\tau D) ,\label{matrix3} \nonumber \\
{\cal L}_{D\bar{D}J/\psi} &=&
ig_{DD\psi} \psi \cdot
{D}\overleftrightarrow{\partial}\bar{D}.
\end{eqnarray}
The three couplings in (\ref{matrix3})
are related to the single parameter $g_2$ as
$\frac{g_{D^*D^*\psi}}{m_{D^*}} = \frac{g_{DD\psi}}{m_D}= 
g_{D^*D\psi}= 2 g_2 \sqrt{m_\psi }$
and $g_2=\frac{\sqrt{m_\psi}}{2m_Df_\psi}$ with $f_\psi=405$ MeV. 

With above Lagrangians, the potential for the interactions can be constructed, which is presented explicitly in~\ref{Sec: potential}.
 
\section{The numerical results}\label{sec3}

The amplitude ${\cal T}$ for the $\pi^-J/\psi-D^-D^{*0}-D^{*-}D^0$ interaction can be obtained 
by  discretizing the momenta ${\rm p}$,
${\rm p}'$, and ${\rm p}''$  in the integral equation~(\ref{Eq: BS_PWA}) by the Gauss quadrature with a weight $w({\rm
p}_i)$. After such treatment, the integral equation can be transformed to a matrix equation  ~\cite{He:2015mja}
\begin{eqnarray}
{T}_{ik}
&=&{V}_{ik}+\sum_{j=0}^N{ V}_{ij}G_j{T}_{jk}.\label{Eq: matrix}
\end{eqnarray}
The propagator $G$ is a diagonal matrix as
\begin{eqnarray}
	G_{j>0}&=&\frac{w({\rm p}''_j){\rm p}''^2_j}{(2\pi)^3}G_0({\rm
	p}''_j), \nonumber\\
G_{j=0}&=&-\frac{i{\rm p}''_o}{32\pi^2 W}+\sum_j
\left[\frac{w({\rm p}_j)}{(2\pi)^3}\frac{ {\rm p}''^2_o}
{2W{({\rm p}''^2_j-{\rm p}''^2_o)}}\right],
\end{eqnarray}
with on-shell momentum
\begin{eqnarray}{\rm p}''_o=\frac{1}{2W}\sqrt{[W^2-(M_1+M_2)^2][W^2-(M_1-M_2)^2]}.\label{Eq: mometum onshell}
\end{eqnarray}

The rescattering amplitude $T$ can be solved as  $T=(1-{ V} G)^{-1}V$.
The  pole of  rescattering amplitude can be found at $|1-VG|=0$ after  analytic continuation total energy $W$
into the complex plane as $z$.  The  amplitude for the $Y(4260)$ decay $M$ can be written as $M=A^{bk}+TGA$ with the on-shell element being chosen.

\begin{figure}[h!]
\includegraphics[bb=120 50 410 300,clip,scale=1.15]{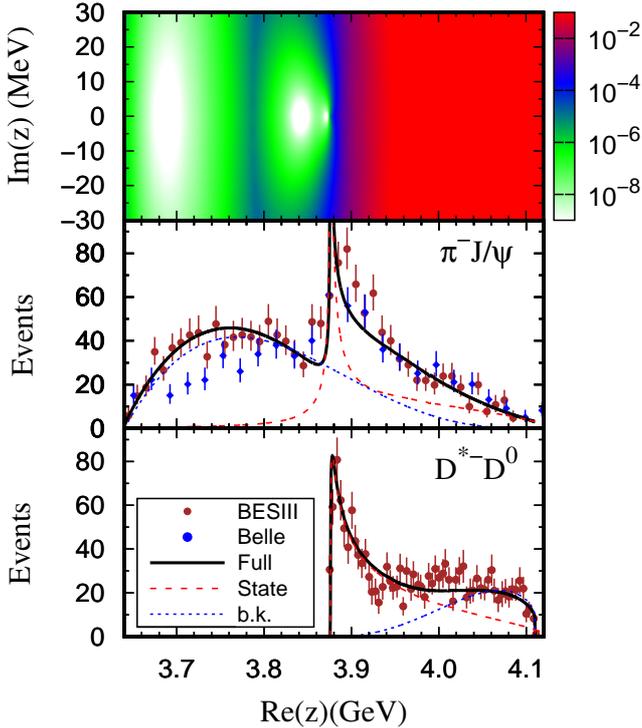}
\caption{ The $\log|1-V(z)G(z)|$ for the the $\pi^-J/\psi-D^-D^{*0}-D^{*-}D^0$ interaction(upper panel). The invariant mass spectra for the $\pi^-J/\psi$ (middle panel) and the $D^{*-}D^0$ with $1^+(1^{+})$ (right panel) are drawn to the same scale. \label{Fig: Invariant mass spectrum}}
\end{figure}

In our model, the parameters in the Lagrangians  are determined by the heavy quark symmetry. The free parameters are the cutoff $\Lambda$ and the $a$, $b$ and $c$ for the background. The cutoffs in the regularization and in the form factor for the exchanged meson have the same value for simplification.  In this work, we try to reproduce the line shapes and relative magnitudes of the $J/\psi\pi^-$ and $D^{*-}D^0$ invariant mass spectra by varying the parameters. When comparing the theoretical results and the experimental data, we should be careful about the number of the events of  two channels which were obtained with different efficiencies in experiments. Fortunately, in original report of  BESIII~\cite{Ablikim:2013xfr}, both the cross sections  and corresponding numbers of the events for the $Z_c(3900)$ in $\pi^-J/\psi$  channel and $Z_c(3885)$ in $D^{*-}D$ channel were presented as 13.5~pb with 307 events and 83.5 pb with 502 events, respectively.  Here we adopt the $M_{max}(\pi^\pm J/\psi)$ distribution as the $J/\pi^-\psi$ invariant mass spectrum to avoid the reflection peak because the background contribution is parametrized  in this work. The theoretical results for events can be  obtained by  multiplied  the efficiencies on theoretical decay distribution for $\pi^-J/\psi$ and $D^{*-}D^0$, respectively. Besides, the different bin sizes adopted in two channels are also considered in the calculation.  After such treatment and a general normalization to the experimental data,  the comparison between the theoretical and experimental results can be carried out. It is found that with a cutoff $\Lambda=0.185$ GeV the invariant mass spectra can be reproduced as shown in Fig.~\ref{Fig: Invariant mass spectrum}, and the corresponding parameters for the background are $(a,b,c)=(0.5,1.2,3.6)$ and $(1.0,0.05,8.5)$ for the $\pi^-J/\psi$ and $D^{*-}D^0$ invaraint mass spectra, respectively.

At low energies, the $\pi^-J/\psi$ invariant mass spectrum is mainly from the background contribution, which decreases with  increase of the energies near and higher than the $D^{*-}D^0$ threshold. A sharp peak arises near the threshold due to the $\pi J/\psi-\bar{D}^*D$ rescattering, which effect decreases a little slower at energies above the threshold than at energies  below the threshold.  The full model can reproduce the  $\pi^-J/\psi$ invariant mass spectrum generally.  The peak seems too sharp compared with the experiment, which may be from the contributions neglected in this work. 
As in our previous work in Ref.~\cite{He:2015mja} where only $D^*\bar{D}$ scattering were considered, the $D^{*-}D^0$ invariant mass spectrum  of the $Y(4260)$ decay can be reproduced. At low energies, the peak near threshold is almost from the $\pi J/\psi-\bar{D}^*D$ rescattering and the background contribution becomes important at higher energies.  Combined the results of both invariant mass spectra,  the $Z_c(3900)$ in $\pi^-J/\psi$ invariant mass spectrum and the $Z_c(3885)$ in  $D^{*-}D^0$ invariant mass spectrum  can be reproduced simultaneously  from the $\pi^-J/\psi-D^{*-}D^0-D^-D^{*0}$ rescattering.

Though peaks can be produced in the invariant mass spectra, we still need  to find out that the resonance structures are from a pole or just  cusps.  In the literatures~\cite{Hyodo:2013iga,Hanhart:2014ssa}, the category of the pole from the two-body interaction has been studied. In the order of the attraction of interaction from strong to weak, there exist four types of poles, bound state which is below threshold and usually called molecular state,  virtual state which is also below the threshold but in the second Riemann surface, virtual  state with width which is below the threshold but has an imaginary part, and resonance which is beyond the threshold and has an imaginary part.  Hence, we adjust the cutoff, with which the strength of the interaction has positive correlation.   The poles produced form the  $\pi^-J/\psi-D^{*-}D^0-D^-D^{*0}$ interaction with typical cutoffs are listed in
Table~\ref{Tab: bound state1}.
\renewcommand\tabcolsep{0.25cm}
\renewcommand{\arraystretch}{1.5}
\begin{table}[h!]
\begin{center}
\caption{The  bound states from the $\pi^-J/\psi-D^{*-}D^0-D^-D^{*0}$  interaction with typical cutoffs $\Lambda$.
The cutoff $\Lambda$ and  energy $W$ are in units of GeV, and MeV, respectively. \label{Tab:
bound state1}
\label{diagrams}}
	\begin{tabular}{cl|cl|cl|cl}\bottomrule[1.5pt]
\multicolumn{4}{c|}{Full model}  &
\multicolumn{4}{c}{No $\pi^-J/\psi$}  \\\hline
 $\Lambda$ &  \multicolumn{1}{c|}{$W$ }  & $\Lambda$ &  \multicolumn{1}{c|}{$W$ }& $\Lambda$ &  \multicolumn{1}{c|}{$W$ } & $\Lambda$ &  \multicolumn{1}{c}{$W$ }   \\\hline
 1.5   & 3833  & 2.1  & 3875+0i   &  1.6   & 3837     & 2.4  & 3875       \\ 
 1.7   & 3862  & 2.3  & 3873+1i   &  1.8   & 3865     & 2.6  & 3873  \\
 1.9   & 3873  & 2.5  & 3863+3i   &  2.0   & 3873     & 2.8  & 3867   \\
 2.0   & 3875  & 2.7  & 3845+5i   &  2.2   & 3875     & 3.0  & 3856  \\  
\toprule[1.5pt]
\end{tabular}
\end{center}

\end{table}

First, we list both the results for the $\pi^-J/\psi-D^{*-}D^0-D^-D^{*0}$ interaction and these after turning off the $\pi^-J/\psi$ channel. From the results, one can find that the interaction is dominant with the $D^*\bar{D}$ interaction. 
If the cutoff larger than about 2.1 GeV, a pole below the $\bar{D}^*D$ threshold is produced from the interaction. Compared with the results without $\pi^-J/\psi$ channel, the imaginary part of the pole is obviously from the coupled-channel effect. By varying the values if cutoff a little, the results in full model and these without $\pi^-J/\psi$ channel are  almost same. Hence, it is a bound state mainly from the  $\bar{D}^*{D}$ interaction. With the decrease of the cutoff, the interaction becomes weaker and the pole is running to the threshold. When the cutoff is smaller than about 2.1 GeV, a pole will appear in the second Riemann surface of the  $\bar{D}^*{D}$ interaction. This pole is leaving the threshold with the decrease of the cutoff and will merge with the lower pole as shown in Fig.~\ref{Fig: Invariant mass spectrum}. If the cutoff decreases further, the pole dies away and no virtual state with width is produced, though a peak  still can be produced as a cusp near the $\bar{D}^*D$ threshold in the invariant mass spectra which is much wider than the experimental $Z_c(3885)$.  Combined with results in  Fig.~\ref{Fig: Invariant mass spectrum} and  in
Table~\ref{Tab: bound state1}, one can find both $Z_c(3900)$ and $Z_c(3885)$ are from a virtual bound state mainly from the  $\bar{D}^*{D}$ interaction. The invariant mass spectra can not be reproduced with a bound state or cusp effect without pole.

\begin{figure}[h!]
\includegraphics[bb=125 50 410 230,clip,scale=1.3]{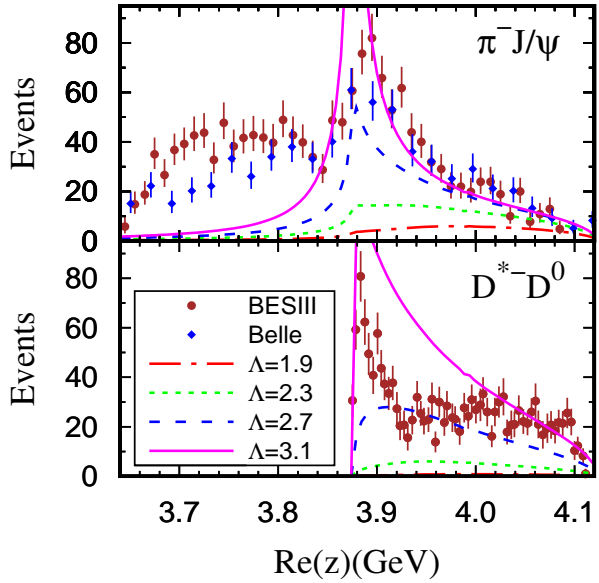}
\caption{ The invariant mass spectra obtained only  with  coupling between $J/\psi\pi^-$ channel and $D^{*-}D^0-D^-D^{*0}$ channel. \label{Fig: noDAD}}
\end{figure}

The above results suggest that the $\pi^-J/\psi$ channel plays a minor role in the $\pi^-J/\psi-D^{*-}D^0-D^-D^{*0}$ interaction, which also leads to a relatively small decay width in $\pi J/\psi$ channel compared with that in $\bar{D}^*D$ channel reported at BESIII~\cite{Ablikim:2013xfr}.  It is interesting to give the results only with coupling between $J/\psi\pi^-$ channel and $D^{*-}D^0-D^-D^{*0}$ channel.  From the results in Fig.~\ref{Fig: noDAD}, if we increase the cutoff to a value about 3 GeV, the peak in the $J/\psi\pi^-$ invariant mass spectrum can be reproduced.  However,  the peak in the $D^{*-}D^0$ invariant mass spectrum is much wider than the experiment. No pole is produced from the interaction, and the peaks are from the cusp effect. Hence, in our model, with only the coupling of $J/\psi\pi^-$ and $D^{*-}D^0-D^-D^{*0}$ channels, the $D^{*-}D^0$ invariant mass spectrum can not be explained. This result supports two structures are from the virtual  state mainly from the  $\bar{D}^*{D}$ interaction.

\section{Summary}\label{sec4}

In this work, the $\pi J/\psi$ and $\bar{D}^*D$ invariant mass spectra of the $Y(4260)$ decay is studied with rescattering of $\pi J/\psi-\bar{D}^{*}D$, which is calculated in a quasipotential Bethe-Salpeter eqaution approach.  The theoretical invariant mass spectra are compared with BESIII experiment to determine the pole structure of $\pi J/\psi-\bar{D}^{*}D$ interaction. 

The peaks in both invariant mass spectra are reproduced from the $\pi J/\psi-\bar{D}^{*}D$  rescattering in the $Y(4260)$ decay.  When the experimental data at BESIII is reproduced, the $\pi J/\psi-\bar{D}^*D$ interaction produce a virtual  state at energy of about 3870 MeV.  The $\bar{D}^{*}D$ channel plays important role to produce the virtual state and the coupling between $\pi J/\psi$  and $\bar{D}^{*}D$  is relatively small, which is consistent with experimentally observed larger cross section of $Z_c(3855)$ in $\bar{D}^*D$ channel than that of $Z_c(3900)$ in the $\pi J/\psi$ channel.
After turning off the $\bar{D}^{*}D$ interaction and keeping only the coupling between $\pi J/\psi$  and $\bar{D}^{*}D$, the cusp effect still can give peaks near the  $\bar{D}^{*}D$ threshold. However, the peak in the $\bar{D}^*D$ invariant mass spectrum is quite broad, which conflicts with the BESIII experiment.  

\vskip 10pt

\noindent {\bf Acknowledgement} Authors thank Dr. Bin Zhong for useful discussion. This project is supported by the National Natural Science
Foundation of China (Grants No. 11675228, No. 11375240, and No. 11775050), and the Major State Basic Research Development
Program in China under grant 2014CB845405.

\appendix

\section{Phase space in the center of mass frame}\label{Sec: phasespace}

To study the invariant mass spectrum of  particles 2 and 3, it is convenient to rewrite the Lorentz-invariant phase space $d\Phi$  in the center-of-mass  frame of particles 2 and 3.  With such treatment, the results of the Bethe-Salpeter equation in the center of mass frame also can be embedded directly. Thus, we first rewrite the phase factor as~\cite{Kamano:2016djv}
\begin{align}
d\Phi&=(2\pi)^4\delta^4(P-p_1-p_2-p_3)d^3\tilde{p}_1d^3\tilde{p_2}\tilde{p}_3\nonumber\\
&=(2\pi)^4\delta(E^{cm}_2+E^{cm}_3-W_{23})\delta^3({\bm p}^{cm}_2+{\bm p}^{cm}_3)d^3\tilde{p}_1d^3\tilde{p}_2^{cm}d^3\tilde{p}_3^{cm} 
\end{align}
where $d\tilde{p}=d^3p/[(2\pi)^32E]$  and $W_{23}^2=(M-E_1)^2-|{\rm p}_1|^2$. Here the Lorentz invariance of the  $d^3\tilde{p}$ and $\delta^4(P-p_1-p_2-p_3)$ is used.

Owing to the three-momentum $\delta$ function, the integral over ${\bm p}^{cm}_2$ can be eliminated.  The momentum of the particle 3 has a relation ${\rm p}^{cm}_3=\frac{1}{2M_{23}}\sqrt{\lambda(M_{23}^2,m_3^2,m_2^2)}$ with  invariant mass of the $23$ system $M_{23}=E^{cm}_2+E^{cm}_3$. Now the quantity $d^3p^{cm}_3$ can be converted to $dM_{23}$ by the relation,
\begin{align}
d^3p^{cm}_3=\frac{E^{cm}_2E^{cm}_3{\rm p}^{cm}_3}{M_{23}}dM_{23}d\Omega^{cm}_3.
\end{align}
The energy-conserving $\delta$ function is substituted as,
\begin{align}
\delta(M_{23}-W_{23})=\frac{W_{23}}{|M{\rm p}_1/E_1|}\delta(\breve{\rm p}_1-{\rm p}_1)
\end{align}
where the $\breve{\rm p}_1$ satisfies $M_{23}^2=(M-\breve{E}_1)^2-\breve{\rm p}_1^2$.
Performing the integral over ${\rm p}_1$, we obtain the final expression of phase space factor,
\begin{align}
d\Phi
&=\frac{1}{(2\pi)^5}\frac{\breve{\rm p}_1{\rm p}^{cm}_3}{M}d\Omega_1d\Omega^{cm}_3dM_{23}.
\end{align}

Now, we treat the three-body amplitude of the decay of $Y(4260)$ with rescattering which is written as 
\begin{align}
&{\cal M}^{Z}_{\lambda_2,\lambda_3;\lambda}(p_1,p_2,p_3)\nonumber\\&=\int \frac{d^4p'_3}{(2\pi)^4} {\cal T}_{\lambda_2,\lambda_3}(p_2,p_3;p'_2,p'_3)  G(p'_3){\cal A}_\lambda(P,p'_2,p'_3).
\end{align}
With the Lorentz invariance,  the amplitude can be rewritten in the center-of-mass frame of particles 2 and 3 as 
\begin{align}
&{\cal M}^{Z}_{\lambda_2,\lambda_3;\lambda}(p_1,p_2,p_3)
=\int \frac{d^4p'^{cm}_3}{(2\pi)^4} {\cal T}_{\lambda_2,\lambda_3}(p^{cm}_2,p^{cm}_3;p'^{cm}_2,p'^{cm}_3) \nonumber\\&\cdot G(p'^{cm}_3){\cal A}_\lambda(P^{cm},p'^{cm}_2,p'^{cm}_3).
\end{align}

After partial-wave decomposition, the amplitude is
\begin{align}
&{\cal M}^{Z}_{\lambda_2,\lambda_3;\lambda}(p_1,p_2,p_3)\nonumber\\&=\sum_{J\lambda_R}N_{J}D^{J*}_{\lambda_R,\lambda_{32}}( \Omega_3^{cm})\sum_{\lambda'_2\lambda'_3}\int \frac{{\rm p}'^{cm2}_3d{\rm p}'^{cm}_3}{(2\pi)^3}\nonumber\\
&~ \cdot ~i{\cal T}^J_{\lambda_2,\lambda_3;\lambda'_2,\lambda'_3}({\rm p}'^{cm}_3) G_0({\rm p}'^{cm}_3) {\cal A}^{J}_{\lambda'_2,\lambda'_3;\lambda}({\rm p}'^{cm}_3,\Omega_1),
\end{align}
The momentum of $Y(4260)$ and final $\pi$ in the center of mass frame of particles 2 and 3 are 
\begin{align}
{\bm P}^{cm}&=\frac{M}{M_{23}}{\bm p}_1,
P^{cm0}=\frac{1}{M_{23}}(M-E_1({\bm p}_1))M;\nonumber\\
{\bm p}^{cm}_1&=\frac{M}{M_{23}}{\bm p}_1,
p_1^{cm0}=\frac{1}{M_{23}}\left[(M-E_1({\bm p}_1))M-M^2_{23}\right].
\end{align}

\section{One-boson-exchange potential}\label{Sec: potential}
Here, we present the explicit form of the one-boson-exchange potential. The potentials for the $D^-D^{*0}\to D^-D^{*0}$ interaction with vector $\mathbb{V}$, $J/\psi$ and $\sigma$ meson exchanges are
\begin{align}
{\cal V}_{\mathbb{V}}&=\frac{iI_\mathbb{V}\beta^2g^2_V}{2(q^2-m_\mathbb{V}^2)}
	(k_2+k'_2)\cdot(k_1+k'_1)\epsilon_2\cdot\epsilon'_2,\nonumber\\
{\cal
	V}_{J/\psi}&=\frac{-ig_{D^*D^*J/\psi}g_{DDJ/\psi}}{q^2-m_{J/\psi}^2}	[\epsilon'_2\cdot(k_1+k'_1)~\epsilon_2\cdot(k_2+k'_2)\nonumber\\
	&+\epsilon'_2\cdot(k_2+k'_2)~\epsilon_2\cdot(k_1+k'_1)
	-(k_2+k'_2)\cdot(k_1+k'_1)~\epsilon'_2\cdot\epsilon_2],\nonumber\\
	{\cal V}_{\sigma}&=\frac{i4g^2_\sigma
m_{P}m_{P^*}}{q^2-m^2_\sigma}\epsilon_2\cdot\epsilon'_2.
\end{align}
 The potential for the $D^{*-}D^{0}\to D^{*-}D^{0}$ interaction can be obtained from these for the $D^-D^{*0}\to D^-D^{*0}$ interaction by alternating particle 1 and particle 2.

The potentials  for the $D^{*-}D^{0}\to D^-D^{*0}$ interaction with  $\mathbb{V}$, $J/\psi$  and $\mathbb{P}$ meson exchanges are
\begin{align}
{\cal V}_\mathbb{V}&=\frac{-i2I_\mathbb{V}\lambda^2
g^2_V}{q^2-m_\mathbb{V}^2}\varepsilon_{\lambda\alpha\beta\mu}(k_2+k'_2)^\lambda
q^\alpha
\epsilon'^\mu_2~\varepsilon_{\lambda'\alpha'\beta\nu}(k_1+k'_1)^{\lambda'}q^{\alpha'}\epsilon^\nu_1,\nonumber\\
{\cal
	V}_{J/\psi}&=\frac{ig_{DD^*J/\psi}^2}{q^2-m_{J/\psi}^2}
	\epsilon^{\beta\mu\alpha\tau }q^\beta\epsilon'^\alpha_2
	(k'_2+k_2)^\tau\epsilon^{\beta'\mu\alpha'\tau'
	}q^{\beta'}\epsilon_1^{\alpha'}
(k'_1+k_1)^{\tau'},	\nonumber\\
	{\cal
	V}_{\mathbb{P}}&
	=-i\frac{4I_\mathbb{P}g^2m_Pm_{P^*}}{f^2_\pi(q^2-m^2_\mathbb{P})}
	q\cdot\epsilon_1 q\cdot\epsilon'_2.
\end{align}
For vector  meson exchange flavor factor $I_{\rho}=-I_{\omega}=1/2$, and for the pseudoscalar meson $I_{\pi}=-3I_{\eta}=-1/2$.

For the coupling of the $D^-D^{*0}\to \pi^-J/\psi$ interaction, there exist two type of  potentials, $t$ and $u$,  as 
\begin{eqnarray}
{\cal
	V}_{D^*, t}&=&
	\frac{-i2g\sqrt{m_Dm_{D^*}}g_{D^*D^*J/\psi}}{f_\pi(q^2-m_{D^*}^2)}k^\mu_1(-g^{\mu\nu}+q^{\mu}q^\nu/m_{D^*}^2)\nonumber\\&\cdot&[\epsilon'_2\cdot\epsilon_2
	(k_2-q)^\nu-\epsilon'_2\cdot(k_2-q)\epsilon_2^\nu+\epsilon'^\nu_2~\epsilon_2\cdot(k_2-q)],
	\nonumber\\
{\cal	V}_{D,u}	&=&
	\frac{-i4g\sqrt{m_Dm_{D^*}}g_{DDJ/\psi}}{f_\pi(q^2-m_D^2)} k'_1\cdot\epsilon_2~k_1\cdot\epsilon'_2,\nonumber\\%
{\cal	V}_{D^*,u}
	&=&
\frac{-4igg_{J/\psi D^*D}}{f(q^2-m_{D^*}^2)}\epsilon_{\alpha\beta\rho\mu}\epsilon^\beta_2q^\alpha k'^\rho_1~\epsilon^{\alpha\beta\nu\tau}
	k'^\alpha_2\epsilon'^\beta_2 q^\tau .
\end{eqnarray}	
Here, $q_{t}=k'_2-k_2$ and $q_{u}=k'_1-k_2=k_1-k'_2$. 
The potential of the  $D^{*-}D^{0}\to \pi^-J/\psi$ interaction can be obtained from these of the $D^-D^{*0}\to \pi^-J/\psi$ interaction by alternating initial particles 1 and 2.

A form factor is introduced to compensate the
off-shell effect of exchanged meson~\cite{Feuster:1996ww,Feuster:1997pq}
\begin{equation}f(q^2)=\frac{\Lambda^4+(q^2_t-m^2)^2/4}{\Lambda^4+(q^2-(q_t^2+m^2)/2)^2},\end{equation}
where $q_t^2$ denotes the value of $q^2$ at the kinematical threshold. The kinematical regime between the threshold and the on-shell point of the exchange particle is stressed and  $t$-channel contributions at threshold are directly given by their couplings. The form factor is only function of the Lorentz invariant $q^2$, pole free on the real $q^2$ axis,  normalized to 1 for $q^2 = m^2$ and $q^2=q^2_t$, but does not have its maximum at $q^2 = m^2$. 
In the propagator of the meson exchange we make a replacement $q^2\to-|q^2|$ to remove the singularities as Ref.~\cite{Gross:2008ps}.

%

%\bibliography{../../../reference/Jabref/bibliography}

\end{document}